# Pulses, Spectral Lags, Durations, and Hardness Ratios in Long GRBs

Jay P. Norris[*], Jeffrey D. Scargle[†], and Jerry T. Bonnell[*¶]

[*]*Laboratory for High Energy Astrophysics,*
*NASA/Goddard Space Flight Center, Greenbelt, MD 20771, USA*

[†]*Space Science Division, NASA/Ames Research Center, Moffett Field, CA 94035-1000, USA*

[¶]*Universities Space Research Association, Washington, DC, USA*

**Abstract.** We analyze BATSE 64-ms data for long gamma-ray bursts ($T_{90} > 2.6$ s), exploring the relationships between spectral lag, duration, and the number of distinct pulses per burst, $N_{pulses}$. We measure durations using a brightness-independent technique. Within a similarly brightness-independent framework, we use a "Bayesian Block" method to find significant valleys and peaks, and thereby identify distinct pulses in bursts. Our results show that, across large dynamic ranges in peak flux and the $N_{pulses}$ measure, bursts have short lags and narrow pulses, while bursts with long lags tend to have just a few significant, wide pulses. There is a tendency for harder bursts to have few pulses, and these hardest bursts have short lags. Spectral lag and duration appear to be nearly independent – even for those bursts with relatively long lags: wider pulses tend to make up in duration for fewer pulses. Our brightness-independent analysis adds to the nascent picture of an intimate connection between pulse width, spectral lag, and peak luminosity for bursts with known redshifts: We infer that lower-luminosity bursts should have fewer episodes of organized emission – wider pulses with longer spectral lags.

## INTRODUCTION

Gamma-ray burst (GRB) time profiles are notoriously heterogeneous – chaotic and unpredictable in appearance – sufficiently so to evade physical modeling attempts. The best demonstration that each GRB time profile is unique and unpredictable comes from burst gravitational lens searches. The results of Marani et al. show that no two BATSE bursts studied have identical temporal and spectral development [1].

One of the first quantitative indications of a global tendency was Nemiroff's "ψ" temporal asymmetry analysis [2], which showed that bursts tend to be asymmetric on all timescales. Even for bursts at one extreme, where spike-like pulses are nearly symmetric at BATSE energies, the burst envelope is often asymmetric. The "Pulse Paradigm" further elucidates burst behavior: Pulses range from narrow and nearly symmetric, to wide and asymmetric, with low energy lagging high energy [3]. Thus, diversity at the level of burst duration masks a determinism at the pulse level: *Individual pulses are organized in time and energy*, and pulses within a given burst tend to exhibit the same degree of asymmetry and spectral lag [3]. In fact, for the few bursts with redshifts there appears to be an important correlation between pulse width, spectral lag, and luminosity [4]. The physical mechanisms giving rise to these correlations are not yet well elucidated. However, recent work by Soderberg and Fenimore shows that a pulse's locus in time in { Intensity, Peak in $\nu \cdot F(\nu)$ } space is inconsistent with pure kinematics of colliding shells, requiring the inclusion of cooling mechanisms to explain pulse profiles [5].

The question we begin to address here is: For a given burst, how is pulse behavior contingent upon global behavior – how are number of pulses, duration, spectral lag, and hardness interrelated?

## ANALYSIS

To better inform studies of pulse behavior, we performed the first automated, brightness-independent analysis of the number of pulses in a burst, $N_{pulses}$, and then examined the relationships of $N_{pulses}$ to hardness ratio, duration, and spectral lag. We used Bayesian Block methods developed by Scargle [6,7] in conjunction with brightness- and noise-equalization methods [8] to estimate $N_{pulses}$ and durations.

The sample comprises 659 long bursts, including all BATSE GRBs with background fits, peak flux (PF) > 1.3 photons cm$^{-2}$ s$^{-1}$, peak intensity (PI) > 4000 counts s$^{-1}$, and $T_{90}$ > 2.6 s, where the duration for each burst was measured after equalization of signal to noise (S/N) and PI to average sample threshold levels. Spectral lags between BATSE channels 1 and 3 (100–300 and 25–50 keV) were measured via the cross correlation function (CCF) [4]. Durations were measured as described in reference [8], also with S/N levels equalized to the average threshold level. The burst data files at 64-ms resolution and their associated background fits are available at the COSSC web site: http://cossc.gsfc.nasa.gov/cossc/batse/index.html .

We bootstrapped the data, making 51 time profiles per burst, fitting a cubic to the CCF, and requiring the fit to be concave up for 51 consecutive trials; else, we increased the length of the fitted interval and restarted the set of fits. The CCF peak was taken as the spectral lag measure, using the measurements of individual fits to generate error bars. We applied the Bayesian Block (BB) cell coalescence method [7] to the S/N equalized time profiles, estimating significant peaks and valleys. The BB representation thus yields the number of identifiable pulse peaks per burst ($N_{pulses}$).

## RESULTS

Figure 1 illustrates two scatter plots of peak flux (50–300 keV) versus spectral lag between BATSE energy channels 1 and 3. The obvious general trend, previously reported for a sample with a higher PF threshold [4,9], is that bursts with longer CCF lags tend to occur at lower peak fluxes. Now, extension of the sample down to PF > 1.3 photons cm$^{-2}$ s$^{-1}$ reveals some bursts with even longer lags, up to ~ 4 s. Notice, however, that most lags tend to concentrate near 0 to ~ 100 ms, even at low PF. For the few GRBs with associated redshifts, luminosity and spectral lag appear to be inversely correlated, following an approximate power law with slope ~ -1 [4,9], and therefore most BATSE bursts with PF _ 1 photons cm$^{-2}$ s$^{-1}$ are probably highly luminous. At lower peak fluxes the situation may be different (see Discussion).

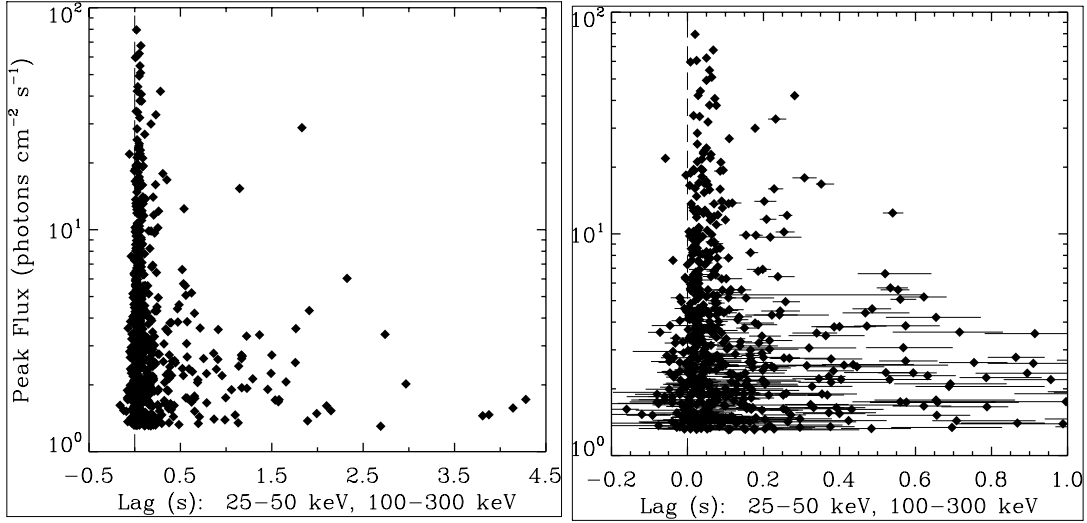

**FIGURE 1.** Peak Flux versus Spectral Lag. Left panel, showing full spectral lag range, illustrates tendency for lower peak-flux bursts to have longer lags. Right panel: Lag scale is magnified, thereby showing only bursts with lags < 1 s. Lag error bars, larger at lower peak flux, indicate that most long lag determinations are significantly different than the majority of lags, which concentrate _ 100 ms.

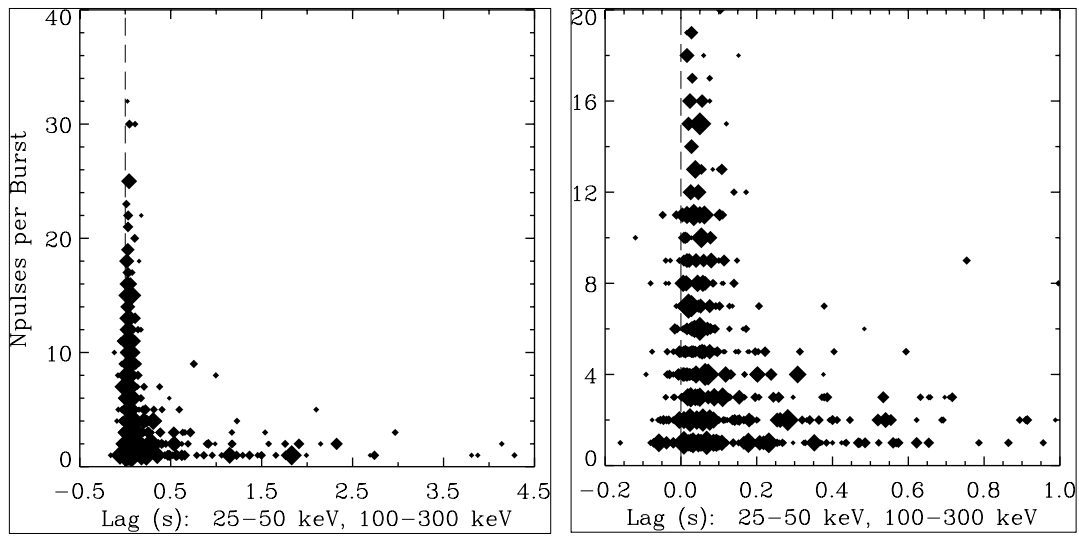

**FIGURE 2.** Number of Pulses versus Spectral Lag. Symbol size is proportional to ln(PF). Left panel: Full ranges of $N_{pulses}$ and spectral lag are shown. Right panel: Lag and $N_{pulses}$ scales are magnified to show detail. Bursts with long lag and few (wide) pulses tend to have low PF (smaller diamonds).

Scatter plots of $N_{pulses}$ versus lag illustrate (Figure 2) an important extension of the "Pulse Paradigm" [3]. While bursts over a wide range in $N_{pulses}$ have short lags, bursts with long lags tend to have just a few significant, *wide* pulses. The connection between luminosity and lag would therefore imply that these long-lag bursts occupy the low-end tail of the *observed* luminosity distribution; their actual numbers must be far greater than observed, due to brightness selection bias. That the pulses in these bursts are wide, monolithic structures – rather than composites of many narrow pulses – is evidenced by the spectral organization revealed in their long lags. To some degree the appearance of bursts with short lags and few pulses must be due to rendering the S/N to threshold level. However, our analysis of relatively bright bursts performed at original S/N levels yields many short lag bursts with few, narrow pulses.

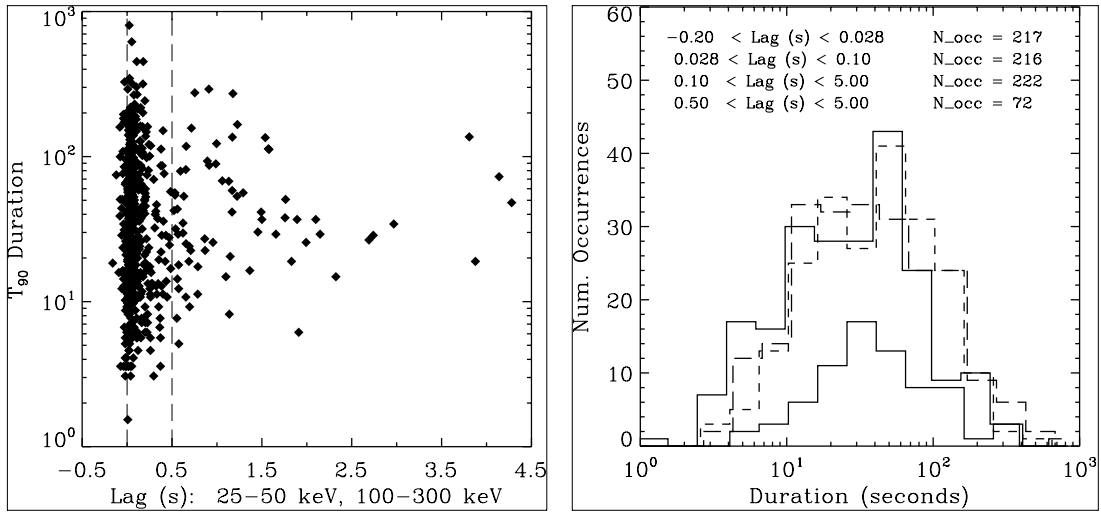

**FIGURE 3.** Left panel: Brightness-independent Duration versus Spectral Lag. Right panel: Duration histograms for three mutually exclusive ranges of lag, which divide the sample nearly equally, and for lag > 0.5 s (that portion of the sample right of dashed line in left panel).

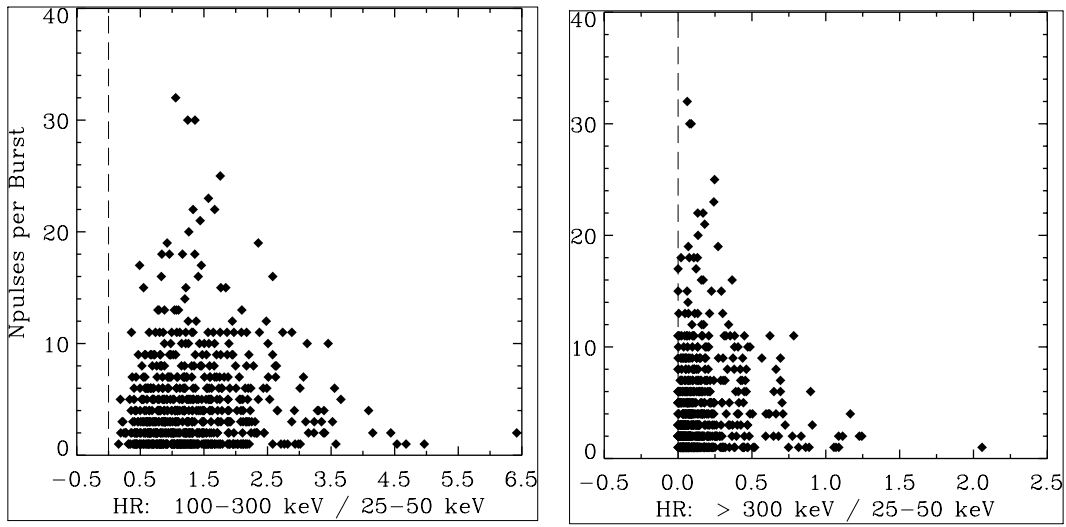

**FIGURE 4.** $N_{pulses}$ versus Hardness Ratio (HR). Left panel: HR = integral counts above background, (100–300 keV) / (25–50 keV). Right panel: Same for (> 300 keV) / (25–50 keV).

The left panel of Figure 3 illustrates a scatter plot of duration versus spectral lag. The right panel shows durations histogrammed for three lag ranges (solid, dotted, dashed – increasing lag) which span the measured values, and for the lags > 0.5 s (lower solid histogram). No trend is apparent; bursts with few long-lag pulses must tend to lengthen the duration by virtue of their wider pulses, as might be expected. Figure 4 illustrates scatter plots of $N_{pulses}$ versus hardness ratio for channels 3/1 ([100–300 keV] / [25–50 keV]) and channels 4/1 ([> 300 keV] / [25–50 keV]). There is a tendency for the harder bursts to have few pulses. Also, the hardest bursts have shorter lags: For HR 3/1 > 2.5, 36 (10) bursts have lags < (>) 60 ms; for HR 4/1 > 0.5, 31 (7) have lags < (>) 60 ms.

## SUMMARY AND DISCUSSION

This work describes the first combined brightness-independent measurements of several parameters for long GRBs ($T_{90}$ > 2.6 s): number of pulses per burst, spectral lag, hardness ratio, and duration. Some interesting relationships are quantified, pertaining mostly to GRBs with few pulses: GRBs with long spectral lag tend to have just a few, wide pulses. At the other extreme, short spectral lag, are GRBs also with few pulses, but with the hardest spectra. Spectral lag and duration appear to be nearly independent – even for GRBs with relatively long lags; wider pulses must compensate in duration for few pulses. From the nascent connection [4,9] between pulse width, spectral lag, and peak luminosity for GRBs with known redshifts we infer that *the lower-luminosity GRBs should tend to have fewer episodes of wide-pulsed emission*. Essentially, wide pulses with long spectral lags means lower luminosity. The analysis of average GRB profiles by Stern et al. as a function of PF [10] suggested an admixture of a larger fraction of simple bursts with wide pulses and long lag near the triggered BATSE PF threshold, ~ 0.25 photon cm$^{-2}$ s$^{-1}$. Since our present sample (PF > 1.3 photon cm$^{-2}$ s$^{-1}$) evidences the beginning of this trend, we infer that low-luminosity, long-lag, few-pulse bursts may dominate near the BATSE PF threshold.

## REFERENCES

1.  Marani, G.F., Nemiroff, R.J., Norris, J.P., Kevin, H., and Bonnell, J.T., *ApJ* **512**, L13 (1999).
2.  Nemiroff, R.J., et al., *ApJ* **423**, 432 (1994).
3.  Norris, J.P., et al., *ApJ* **459**, 393 (1996).
4.  Norris, J.P., Marani, G.F., and Bonnell, J.T., *ApJ* **534**, 248 (2000).
5.  Soderberg, A.M., and Fenimore, E.E., "The Unique Signature of Shell Curvature in Gamma-Ray Bursts," in *Proc. 2$^{nd}$ Rome Workshop on GRBs in the Afterglow Era* (2001).
6.  Scargle, J.D., *ApJ* **504**, 405 (1998).
7.  Scargle, J.D., "Bayesian Blocks: Divide and Conquer, MCMC, and Cell Coalescence Approaches," in *MaxEnt99* (2001).
8.  Bonnell, J.T., Norris, J.P., Nemiroff, R.J., and Scargle, J.D., *ApJ* **490**, 79 (1997).
9.  Norris, J.P., Marani, G.F., and Bonnell, J.T., "Connection between Spectral Lags and Peak Luminosity in GRBs," in *Gamma-Ray Bursts*, edited by R.M. Kippen, R.S. Mallozzi, and G.J. Fishman, AIP Conference Proceedings 526, New York, 2000, p. 87.
10. Stern, B., Poutanen, J., and Svensson, R., *ApJ* **510**, 312 (1999).